\documentclass[openacc]{rsproca_new}

\usepackage{siunitx}

\begin{document}

\title{On the force of vertical winds in the upper atmosphere - consequences for small biological particles}

\author{
A. Berera$^{1}$ and D. J. Brener$^{1}$}

\address{$^{1}$ The Higgs Centre for Theoretical Physics, James Clerk Maxwell Building, Peter Guthrie Tait Road, EH9 3FD, Edinburgh.}

\subject{Geophysics, Atmospheric science, Meteorology}


\corres{Daniel J. Brener\\
\email{danielbrener@ed.ac.uk}}

\begin{abstract}
For many decades vertical winds have been observed at high altitudes of the Earth's atmosphere, in the mesosphere and thermosphere layers. These observations have been used with a simple one dimensional model to make estimates of possible altitude climbs by biologically sized particles deeper into the thermosphere, in the rare occurrence where such a particle has been propelled to these altitudes. A particle transport mechanism is suggested from the literature on auroral arcs, indicating that an altitude of 120 km could be reached by a nanometer sized particle which is higher than the measured 77 km limit on the biosphere. Vertical wind observations in the upper mesophere and lower thermosphere are challenging to make and so we suggest that particles could reach altitudes greater than 120 km, depending on the magnitude of the vertical wind. Applications of the larger vertical winds in the upper atmosphere to astrobiology and climate science are explored.
\end{abstract}

\maketitle

\section{\label{sec:level1}Introduction\protect\\ }

Vertical winds up to 100 m/s have been observed in the upper mesosphere and thermosphere layers of the Earth's atmosphere for many decades. These vertical winds are observed to be sustained on the order of minutes to an hour. High altitude measurements of vertical winds have limited temporal, spatial, and vertical coverage. Ground based measurement techniques using optical imaging are restricted to narrow altitude ranges but have a greater temporal coverage. In situ measurements, using rockets to disperse trackable tracer gases, are temporally limited, but are able to sample larger altitude ranges. Reviews of vertical winds in the thermosphere are given by Smith in 1998 \cite{SMITH19981425} and Larsen and Meriwether in 2012 \cite{Larsen2012}.

Due to the Earth's geomagnetic field, particles in the solar wind are channeled such that they penetrate the atmosphere in the polar regions. The charged particles collide with the neutral air, transferring their energy and momentum. The transfer of momentum and subsequent heating at these altitudes is understood to cause some of the largest observed vertical winds. The mean vertical wind in the polar regions at these altitudes is typically in order of a few tens of m/s. However, upward vertical winds of up to 150 m/s have been observed \cite{Rees1984} \cite{Price1995} \cite{Guo2003} at around 240 km. At polar regions where the aurora occur, known as the auroral zone, vertical winds have been observed with magnitudes around 50 m/s \cite{Wardill1986} \cite{Crickmore1993} \cite{Crickmore1991} \cite{Aruliah1995} \cite{Ishii2001} \cite{Greet2002} \cite{CONDE1995589} \cite{Anderson2012} \cite{ronksley} \cite{Oyama}. A couple of studies have reported vertical winds $>$100 m/s \cite{Rees1984} \cite{Price1995} \cite{Anderson2011}. Most of these observations are of vertical winds at altitudes around 250 km, but some are as low as 90 km (e.g. \cite{Oyama}) and they are usually seen during geomagnetic storms. It must be emphasised that vertical winds in the mesosphere of more than \SI{100}{m/s} have not been observed so far. In Table \ref{tab:winds} we have included a summary of key studies reporting large vertical winds at different altitudes.

In the last year a new set of observations have reported a variance of vertical velocities as high as \SI{60}{m/s} in the mesosphere \cite{chau2021}. Although these values are smaller than those in the thermosphere, they are more than 5$\sigma$ larger than normal vertical velocities variability. These extreme vertical winds were observed when the ionosphere was quiet for a few hours and at an altitude too low for plasma instabilities to be generated. This led the authors to suggest that the origins of the winds were not related to auroral activity. There are also new observations that show unexpected signatures of metallic ion layers at altitudes as high as 180-200 km \cite{chu2021}. These observations indicate that non-auroral activity might be involved in transport of particles deeper into the thermosphere in combination of both neutral vertical winds and geomagnetic Lorentz forcing \cite{chu2021}.

Little theoretical progress has been made in understanding the large vertical wind observations \cite{Larsen2012} \cite{Griffin2018} \cite{Priv_Comm_kath}. Deviations from the hydrostatic balance have been studied (e.g. \cite{Deng2008}, \cite{Zhu2017} \cite{Zhu2020}), but they do not last long enough to create vertical winds sustained over several hours \cite{Larsen2012}. Fully non-hydrostatical models (e.g. the Global Ionosphere Thermosphere Model \cite{Ridley2006}) are unable to reproduce the large vertical winds because they typically run at too low a spatial and temporal resolution \cite{Priv_Comm_kath} \cite{Priv_Comm_Jia}. This is not to say that the vertical winds in the current operational models are wholly unrealistic, but rather that there exists an observational literature of larger vertical winds which have yet to be fully captured by the models (see review paper of \cite{Larsen2012} which came after \cite{Deng2008}). The physical mechanism generating the vertical winds is outwith the scope of this paper, all that matters is the observations of such strong vertical winds.

Various measurements have shown there are particles of a radius around the size of a micron and in reported concentrations of approximately 1 particle cm$^{-3}$ in the stratosphere. \cite{Heintzenberg} \cite{rosen} \cite{ursem} \cite{XU2003201} \cite{yinyan}. Additional studies found that these particles include bacteria \cite{Griffin2004} \cite{ursem} \cite{Wainwright2003}. The highest altitude that biological particles have been found is 77 km \cite{Imshenetsky1978}. These were fungal spores, with a radius within an order of magnitude of a micron. Bacteria have been found as high as 41 km \cite{ursem} \cite{Wainwright2003}. However, these are likely to be underestimates, as the studies were very limited due to the technical difficulty of growing cultures of the bacteria and fungi captured in sealed capsules from rockets sent to these altitudes. 

More recently, cosmic dust samples from the surface of the International Space Station (ISS) were found to have DNA from several kinds of bacteria which were genetically similar to those found in the Barents and Kara seas' coastal zones \cite{ISS_DNA}. The investigators hypothesised that the wild land and marine bacteria DNA could transfer from the lower atmosphere into the ionosphere-thermosphere using the ascending branch of the global electric circuit or the bacteria found may have had a space origin. This paper highlights that there is a body of evidence of large vertical winds in the upper atmosphere which have the capability to push larger particles higher, hence potentially extending the biosphere further than previously thought. Such events are likely to be fluctuations and therefore rare occurrences, but over millennia, may have astrobiological significance.

We are aware that the idea that such biological particles may exist in the upper mesosphere, where the density is usually insufficient to support them for long, is slightly contested in private discussions. However no studies have been conducted that provide alternative explanations for the observations. For our work, it is sufficient to say that such occurrences are probably very rare, but on geological timescales may be significant.


A mechanism known as gravito-photophoresis, arising from irradiation of particles by sunlight, has been shown to elevate micron scale
particles to altitudes of approximately 83 km \cite{Rohatschek1996}. This mechanism has been little investigated with only a handful of papers examining its effects on the upper atmosphere (e.g. \cite{Rohatschek1996} \cite{Horvath2014}).

The existence of noctilucent clouds at altitudes of 80 - 100km provides evidence that small dust-like particles can be found there \cite{Wallace}. These clouds are most often observed closer to the poles or latitudes greater than 50 degrees, and so are also known as polar mesospheric clouds. A definite source of the condensation nuclei is interplanetary space dust from meteors and passing comets. These are known to release large quantities of dust into the upper atmosphere \cite{2006JASTP..68..715R} \cite{plane2012} \cite{wilms}.

However, it is also possible that the dust particles are terrestrial in origin. Volcanic eruptions are terrestrial events capable of significant upward thrust, projecting ash into the stratosphere \cite{tupper} \cite{wilson}.  Modelling of the 1883 Krakatao eruption would indicate that dust from the volcanic ash cloud diffused up to around 85 km \cite{self} \cite{VERBEEK1884}. This idea is supported by observations of noctilucent clouds that appeared at the time of the eruption \cite{ludlam}. A recent study found that 50 - 100 nm sized particles could be projected in volcanic eruptions to the upper mesosphere, and 10 nm to more than 120 km \cite{electrostatic_volcano}. Furthermore, with increasing numbers of commercial spaceflights, especially now including space-tourism, one would reasonably expect to see increases in the incidences of short-lived biological contamination of the mesosphere-lower thermosphere \cite{WEBBER2013138} \cite{astrobio_smith}.


There is a growing interest in the upper atmosphere from the astrobiology community, but there has been no concerted research effort placed on the mesosphere and thermosphere \cite{astrobio_smith} \cite{dassarma_antunes_dassarma_2020}. The recent investigations and publicity around the atmosphere of Venus have also instilled interest in the possibility of life transfer from and to nearby planets \cite{Greaves2020} \cite{venus_debunked} \cite{Hallsworth2021}.

In the last decade, work has been done to extend global numerical weather prediction models into the thermosphere (e.g. WACCM in the US \cite{WACCM}, CMAM in Canada \cite{CMAM} and the Met Office Unified Model in the UK \cite{Griffin2018}) and whole atmosphere modelling studies are considering the role of the mesosphere and thermosphere in the Earth's climate \cite{solomon_15} \cite{thermo_climate} \cite{thermo_climate_2}. Although these models are non-hydrostatic, which allows for the vertical acceleration necessary for vertical winds, there does not presently exist successful modelling studies of the larger vertical winds reported in the observational review papers.

Horizontal winds are, however, well captured by these models. Hence, particle transport in the horizontal is well understood through modelling of Lagrangian coherent structures (e.g. \cite{LCS}), but these studies assume that the vertical motion is \textit{negligible}. Due to the strengths of the vertical winds needed to suspend or propel a large heavy particle upwards, we were motivated to take an unconventional, yet simpler approach to making some first estimates of the maximum altitude attainable by such a particle.

The purpose of this paper is to illustrate the approximate strength of these extreme winds, which has implications for transporting particles larger than molecules higher than experiments have yet measured. We recognise that the natural horizontal symmetry of upper atmospheric particle transport can be broken by the observed large fluctuations in vertical winds, allowing for the complicated horizontal dynamics to be ignored to a good approximation during such events. This provides a clear justification to use a one-dimensional model.

The technicalities of this assumption are left for later discussion in this paper, but in summary, large vertical winds have been correlated with horizontal winds in auroral arcs which may enable a particle to remain in a strong vertical wind flow. This is just one mechanism, and there are likely to be other such processes working at smaller scales. We hope that this short proof-of-concept paper will act as a precursor to future studies using sophisticated 3-D global numerical weather prediction models to reproduce the vertical winds. By considering the possibility of larger particles existing at altitudes higher than previously considered we open up interesting applications in other fields such as astrobiology and climate science which we discuss in the penultimate section.

This paper is presenting a theoretical argument for the presence of heavier nanometer or larger sized particles in the mesosphere and lower thermosphere. To date no dedicated field campaigns have been conducted to search for such particles at these altitudes. By combining observations of strong vertical winds at these altitudes with some theoretical arguments, we show that there is a strong case for the presence of such particles at these high altitudes at intervals dependant on the frequency of strong vertical wind events.

\begin{table}[htb!]
\begin{tabular}{c|c|c|c}
z/km                 & range of maximum w/(m/s) & duration/minutes & e.g.           \\ \hline
240 (F-region)       & 138 - 150              & 15 - 120           & \cite{Rees1984}, \cite{Price1995}, \cite{Guo2003}, \cite{Crickmore1993} \\
120 - 130 (E-region) & 30 - 42                & 15 - 25            & \cite{Larsen2012} \cite{Price1995} \cite{article}             \\
\textless{}110       & 10 - 32                & 20               & \cite{Price1991}              \\
103                 & 10 - 50                & 30               & \cite{Oyama}              \\
86                   & 65                     & 20               & \cite{chau2021}            
\end{tabular}
\caption{\label{tab:winds}Summary of extreme vertical wind observations variation with altitudes of interest, and example supporting studies. Some durations are rough estimates from the figures in the original papers.}
\end{table}


\section{\label{sec:level1}Theory\protect\\ }

\subsection{\label{sec:level2}1-dimensional model}

Empirically speaking, the simplest approximation for the upper atmosphere circulation pattern (i.e. zonal and meridional mean flow) is that these are roughly the same over large regions of the Earth, in magnitude and variation, driven by tides and waves. 
Due to solar particle interactions, the poles are more exceptional, but the approximation remains largely true to zeroth order. Hence, as has been shown in thermosphere tracer particle transport studies, horizontal winds will move a particle considerably around the globe \cite{LCS}. However, as has been noted from vertical wind observations around the world (see review \cite{Larsen2012}), the vertical forces a particle experiences from vertical winds remain similar.

By modelling the particle transport in only the vertical direction, our model assumes that the profile of the vertical wind at different horizontal regions above the Earth is to a good approximation similar.  This can be seen in the observational literature.  Though taken at different locations over the Earth, the vertical wind patterns are approximately the same, with periods on orders of hours of upward and then downward motion and with velocities ranging up to tens of meters per second \cite{Larsen2012}.

Thus a test particle may be blown around considerably in the horizontal direction but in whatever region it is in, the vertical component it experiences has a similar behavior, but weaker strength.  If one were just interested in the vertical motion of this test particle, its dynamics looks somewhat like a random walk, in that it initially may experience upward vertical wind, and then be blown horizontally to a region where it may feel either upward or downward wind and so forth.

If all one is interested in is the vertical motion of the test particle, independent of where it lies horizontally, then one could describe its vertical motion through simply a one-dimensional differential equation. The similarity of the vertical wind over different horizontal regions, implies an approximate horizontal translational symmetry that can be factored out of the motion.  This provides a significant simplification of the problem.

The resulting one-dimensional equation should then be stochastic, with the randomness describing the instances when the test particle feels upward versus downward wind as well as variations in wind speed.  Modelling such an equation as yet is too big a first step.  As a first step, we would just like to assess the strength of vertical wind forces on a test particle and how high such a test particle might traverse.  With this in mind, the first most basic question is how strong a vertical wind is needed to overcome the downward gravitational force.  Here we will develop a simple differential equation to express this modest goal. The equation itself is not new, and has been applied in a different context, as it simply expresses for a directed flow, the terminal velocity a test particle can achieve.  However this equation provides a starting basis for further development, whereby it can be extended into a stochastic equation with more detailed description of drag forces.

For example, our analysis in the next section will go beyond just this equation and include specific cases for auroral arcs over the polar regions. Such details might eventually be understood on aggregate and compactly expressed within a single stochastic equation, but at this early stage, we have little quantitative understanding of how to model all such probabilistic events. This is the first paper to identify this process of nano- to micrometer test particles being propelled upward by vertical wind, and we are simply exploring the physical attributes of this process.

\subsubsection{Continuum approximation}

We will consider test particles climbing from the upper mesosphere, at around 90 km, into the thermosphere. At these high altitudes, the density of the air is 5 orders of magnitude lower than at the surface. In most situations, 3D particle transport is characterised by the ratio of the fluids mean free path to the radius of the test particle that is to be transported, known as the Knudsen number (Kn). The equilibrium mean free path of the air at 90 km is approximately \SI{0.02}{m}, so for a micrometre sized test particle, Kn is 23700 \cite{rubberbible}. Being much greater than unity, this implies that the dynamics of such a test particle are governed by the non-continuum, kinetic regime. This can also be noted by calculating the Reynolds and Mach numbers for the flow which are dominated by the very low densities at these altitudes.

However, in the case of a strong directed 1D flow, such as what has been observed of the vertical winds at these altitudes, to zeroth order the forces in the continuum limit can be applied. We approximate the situation as an entirely vertical directed flow, analogous to an atmospheric river. This point of view is motivated by the fact that the observations show these long lasting streams of upward flow. Effectively, we neglect the kinetic fluctuations which are what dominate the free molecular flow/kinetic regime, characterised by Kn $\ll1$.

If we were to include such fluctuations it would be necessary to consider the mechanics as a 3-dimensional Brownian movement within the directed flow, which is unnecessary detail for this papers objective. We note that there is a substantial body of literature which discusses the various limits of applicability of the continuum approach at high Kn number (e.g. see \cite{doi:10.1063/1.2393436}). It is also worth noting that random lateral fluctuations will be on the scale of the mean free path, and thus smaller than the transport distance of the bulk, i.e. the horizontal winds are much more significant.

Furthermore, in engineering where this problem is encountered for microchannels that have a large aspect ratio (width-to-height), it is conventional to use spanwise space averaging to define an averaged velocity profile, hence defining the equivalent
macroscopic quantities \cite{Karniadakis2005}. This is equivalent to our use of the vertical velocity principally because we neglect in these simple calculations the horizontal motions.

\subsubsection{Threshold velocity for a test particle to ascend with vertical wind}
Our study will focus on a fiducial test particle that has a characteristic size and mass larger than the surrounding constituents, i.e. atoms and molecules, of the atmosphere at that altitude. In describing how a test particle is carried by the wind, two forces shall be considered. The first is the weight of the test particle due to its mass. The second is the force carried by the momentum of the vertical wind, distributed over the surface of the test particle. Using this picture, we obtain the force of a vertical wind acting on the particle. The form of this force is unsurprisingly the same as the standard drag force on a particle falling at terminal velocity, only now acting in the opposite direction to propel the particle.


We model the test particle as a disk with a drag coefficient of unity, for simplicity. This means that all the upward vertical wind hits the test particle at the same time. No significant part of the test particle feels more wind against itself, thereby reducing the problem to one dimension. It is also helpful as it mirrors the most abundant measurements of vertical winds in the upper atmosphere as instantaneous winds at discrete altitude points.

\begin{figure}[!htb]
    \centering
    \includegraphics[scale=0.3]{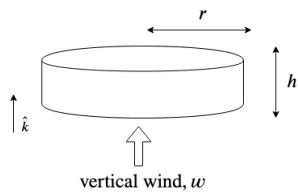}
    \caption{Test particle as a disc, orientated such that the vertical wind, a continuous directed flow, impacts on its major axis, pushing it upwards.}
    \label{fig:diagram_disk}
\end{figure}

Let the test particle be a disc of radius $r$ and thickness $h$. As illustrated in figure \ref{fig:diagram_disk}, and let it sit with its major axis perpendicular to the vertical wind direction. So the density of such a test particle, $\rho_{p}$ can be written as
\begin{equation}
    \rho_{p} = \frac{m}{\pi r^2 h},
    \label{mass_particle}
\end{equation}
where $m$ is the mass, $r$ the radius and $h$ the thickness of the test particle. Consider a vertical wind blowing upwards with velocity, $w$ which impacts on the circular disc surface. Then the mass of the air which will impact on the base surface of the test particle in some finite time $t$ is,
\begin{equation}
    M_{\text{air}} = \rho\pi r^2 w t,
    \label{mass_air}
\end{equation}
where $\rho$ is the density of the air in a cylindrical volume element.

The force due to the vertical wind is essentially the same as that due to the continuum drag force, but acting in the opposite direction, so the test particle can be considered to be moving \textit{with} the flow of the wind after some finite time $t$, to allow the wind to have sufficient force on the test particle. Thus air resistance, against the direction of the test particle's motion, shall be henceforth neglected.

Let the velocity of the vertical wind, $\Vec{w}$ be upwards, perpendicular to the Earth's surface ($\hat{k}$) so that $\Vec{w} = w\hat{k}$. Assume that the vertical wind is constant with both altitude and time. This is a good approximation over some period of time usually lasting tens of minutes and over some
distance extending over several kilometers, when there is a constant vertical wind. Let the velocity of the particle moving upwards with the wind be $\Vec{v} = v\hat{k}$. Hence the velocity of the wind relative to the particle is $(w - v)\hat{k}$. Since all the motion is restricted to one dimension, the $\hat{k}$ unit vectors shall be dropped from now on, with upwards defined as positive.

From equation \ref{mass_air}, the mass of the air impacting on the lower surface of the test particle in some time $t$ is 
\begin{equation}
    M_{\text{air}} = \rho\pi r^2 (w - v) t,
\end{equation}
which gives the momentum transferred by the wind to the test particle,
\begin{equation}
    p = \rho\pi r^2 (w - v)^2 t.
\end{equation}
Hence the force on the test particle due to the wind is given by
\begin{align}
    F_{\text{wind-particle}} &= \frac{dp}{dt} \\
     &= \rho\pi r^2 (w - v)^2.
\end{align}

The equation for the two forces, the weight and vertical wind momentum, is therefore
\begin{equation}
    m\frac{dv}{dt} = -mg + \rho \pi r^2 (w - v)^2. \label{ode_pen}
\end{equation}
As mentioned earlier, this equation is equivalent to the equation for the terminal velocity of a test particle with an opposing drag force proportional to the quadratic vertical wind speed, where the drag coefficient $C_d$ in the quadratic drag law is unity. This occurs when the entire wind flow onto the bottom of the disk comes to rest, creating stagnation pressure.

As the relative velocity force term in the above equation is squared, the force due to the vertical wind cannot change sign. So in the regime where the vertical wind blows downwards, meaning $\Vec{w}<\Vec{v}$, the equation at the moment wrongly suggests the test particle will still move 
upwards. Put simply, taking the limit where $\Vec{w}\rightarrow -\infty$, the acceleration of the test particle will remain in the positive $\hat{k}$ upward direction. To give the force the correct sign, the Heaviside unit step function, $H$ is introduced as
\begin{equation*}
    H(w)=
    \begin{cases}
    1, & \text{if}\ w > 0  \\
    -1, & \text{if}\ w < 0
    \end{cases}
\end{equation*}
Using equation \ref{mass_particle}, and introducing the Heaviside step as above, equation \ref{ode_pen} can be written,
\begin{equation}
        \frac{dv}{dt} = -g + H(w-v) \frac{\rho(z,t)}{\rho_p h} (w(z,t) - v(t))^2.
        \label{ODE}
\end{equation}
Physically this equation says that if the force due to the vertical upward wind is strong enough, it can overcome the force due to the test particle's weight, and cause the test particle to accelerate upwards. This equation is separable, yielding
\begin{equation}
    t = \frac{1}{g} \int^{v(t)}_0 \frac{dv^\prime}{b(w-v^\prime)^2 -1}
    \label{integral_t}
\end{equation}
with the boundary conditions that in some time $t$ the test particle reaches a velocity $v(t)$ having been at rest initially and $b = \frac{H\rho(z)}{gh\rho_p}$. Solving this integral by substitution, and then rearranging, one finds that
\begin{equation}
    v = w - \sqrt{\frac{g}{\lambda }} \left (\frac{1+\left(\frac{\sqrt{\frac{\lambda }{g}}w - 1}{\sqrt{\frac{\lambda }{g}}w+1}\right) \exp(-2t\sqrt{\lambda g})}{1-\left(\frac{\sqrt{\frac{\lambda }{g}}w - 1}{\sqrt{\frac{\lambda }{g}}w+1}\right) \exp(-2t\sqrt{\lambda g})} \right ), \label{full_sol}
\end{equation}
$\text{ where } \lambda = H(w-v) \frac{\rho(z,t)}{\rho_p h}.$
It is assumed that $t$ is sufficiently small such that the air density $\rho(z)$ and vertical wind speed $w$ remain constant. In the limit that $t\rightarrow \infty$, we obtain the correct steady state solution, as found in equation \ref{steady_state}. When $t=0$, we find $v=0$ as expected. From this equation, one finds that the timescale to reach terminal velocity is approximately $\sqrt{\frac{1}{\lambda g}}$. For a standard test particle we consider (see section 2b), this comes out at less than a second, which is shorter than the timescale of the observed large vertical winds, hence enough acceleration can be provided.

The Heaviside function is introduced to maintain some stability when working with real observations which can sometimes have cases where $w<0$, but really this equation is only valid for the cases where $v<w$ and $w>0$.

Let us imagine the situation where initially the test particle is at rest, and then some upward wind blows against it. Provided the wind is stronger than the test particle's weight in the air, the test particle will start to move upwards, accelerating in some finite time $t$ to reach the steady state. This is when the wind just balances the weight of the test particle, and at this point the test particle's velocity will approach, but generally not reach, the speed of the wind. To see this, consider the limiting cases of the parameter $\lambda$ when the wind is blowing upwards such that $H(w-v)$ is positive in equation \ref{full_sol}. When the parameter $\lambda$ is maximised the test particle's velocity is closer to that of the wind. 

Due to the disk geometry chosen, $\lambda$ is maximised when the test particle thickness $h$ is minimised with respect to the density of the test particle $\rho_p$. In other words the optimum shape is that of a pancake where the test particle's mass is distributed over as large a surface area as possible. Additionally, the higher the air density, the stronger the force of the vertical wind, hence the smaller the difference between the speed of the test particle and the speed of the wind. Note that this discussion is only valid when considering directed flow which is the continuum limit to zeroth order when neglecting lateral motions. These considerations are also complimentary to discussion of a continuum drag coefficient.

The steady state solution for equation \ref{ODE}, is $\frac{dv}{dt} = 0$. This corresponds to,
\begin{equation}
    -g + \lambda (w(z,t) - v(t))^2 = 0, \text{ where } \lambda = H(w-v) \frac{\rho(z,t)}{\rho_p h}
\end{equation}
giving in the steady state,
\begin{equation}
    v= w - \sqrt{\frac{g}{\lambda}}. \label{steady_state}
\end{equation}
The negative square-root is selected as $v<w$ is required, and for $v>0$ it follows that $\sqrt{\frac{g}{\lambda}} < w$. This gives the minimum wind needed to get the test particle to reach the steady state,
\begin{align}
    w \geq \left ( \frac{g \rho_p h}{\rho(z)} \right )^{\frac{1}{2}}. \label{terminal}
\end{align}
This condition is trivially the requirement that for particles to move upward, the upward velocity must exceed the terminal velocity. We will refer to this as the threshold velocity.

\subsection{\label{sec:level2}Estimates from vertical wind observations}

We will examine three different test particles to illustrate the forces of vertical winds in the literature and make further hypotheses. The first test particle we will call our \textit{standard test particle}, with general dimension of a nanometer (height and diameter) and standard dust density \SI{1}{g/cm^{3}} \cite{WHO}, giving it a mass of $\sim$\SI{3}{\times10^{-24}kg}.

\begin{figure}[!htb]
    \centering
    \includegraphics[scale=0.5]{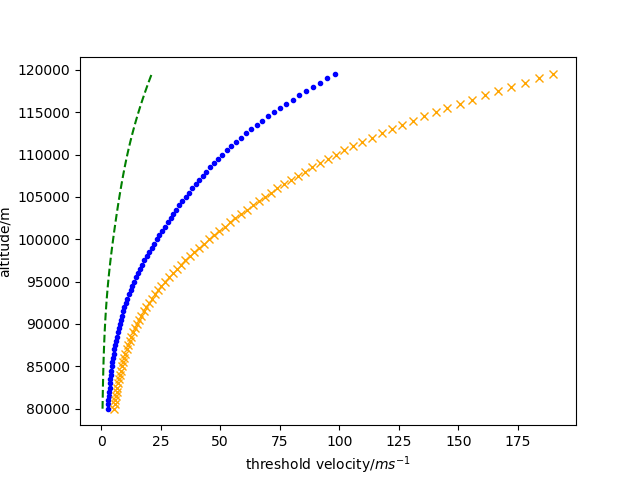}
    \caption{Threshold velocity equation \ref{terminal} for three different test particles. Standard dust test particle of density \SI{1000}{kgm^{-3}}, height and radius of a nanometer with a mass of $\sim$\SI{3}{\times10^{-24}kg} (green dash). Virus sized test particle of density \SI{196}{kgm^{-3}}, thickness \SI{109}{nm} (H1N1 virus from \cite{Wang16028}) (blue dot). Small bacteria or bacteria organelle sized test particle of density \SI{2000}{kgm^{-3}}, height of 40 nm, radius $\sim$\SI{2}{\mu m} and mass of \SI{10}{^{-15}kg} (orange cross).}
    \label{fig:terminal_vel}
\end{figure}

In figure \ref{fig:terminal_vel}, the condition equation \ref{terminal} for our standard test particle and two biologically defined ones is presented. We used the NRLMSISE-00 model (see \cite{2018cosp...42E.893D} and \cite{Picone2002}) by implementing the Python module \texttt{fluids} and its class \texttt{atmosphere.ATMOSPHERE\_NRLMSISE00}, to give the average air density, $\rho(z)$ as a function of altitude \cite{fluids}. The change in the vertical velocity required for such a test particle to reach steady state is exponential and then linear. This corresponds to the change in the air density profile between the mesopause and thermosphere.

In our model, after some short transient period, the velocity of the test particle will become close to that of the wind, which is the threshold velocity. We now use steady state solution, equation \ref{steady_state}, to make some simple estimates for the vertical distance a test particle could be carried upwards in the vertical winds reported by different observational studies.

Measurements of ion velocities can be considered as a proxy for neutral winds below altitudes of around 105 km. In \cite{Oyama}, they used this technique to measure vertical winds in Greenland on two different nights at an altitude of about 103 km. As found in most other vertical wind data, the wind displayed oscillatory behaviour where it switched between upward and downward. The vertical winds range in magnitude between 10 - 50 m/s and on both nights there are long periods of consistent upward wind. If we consider a wind of 50 m/s for our small bacteria or bacteria organelle sized test particle (density \SI{2000}{kgm^{-3}}, height of 40 nm, radius $\sim$\SI{2}{\mu m}, mass of \SI{10}{^{-15}kg}) and use the U.S Standard Atmosphere value for the air density at 100 km of \SI{5.604}{\times 10 ^{-7} kgm^{-3}} \cite{rubberbible}, we find that the upward velocity of the test particle is \SI{13}{m/s} (note there will be a difference from the orange crossed curve in Figure \ref{fig:terminal_vel} as that uses a slightly different air density). Since the wind typically grows to a maximum over around 20-30 minutes lets suppose then the test particle on average has a vertical velocity of $\sim$\SI{7}{m/s}. This would mean that the test particle would climb around \SI{8.4}{km} in 20 minutes.

\begin{figure}
    \centering
    \includegraphics[scale=0.4]{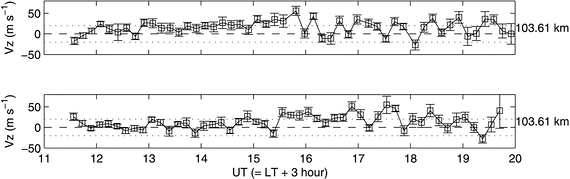}
    \caption{Time series of vertical ion velocities (neutral vertical wind proxy) at 103 km on September 5 (top) and 12 (bottom) 2003. The dotted lines indicate 20 m/s. From \cite{Oyama}. }
    \label{fig:oyama}
\end{figure}

The use of a disc geometry for the test particle means that the greatest forces on the test particle will occur when it maximises its surface area and minimises its thickness with respect to its mass. The stronger the vertical winds, the larger the test particle can be. The approximation is sufficient to within an order of magnitude. However, the most significant approximation in our analysis is the neglecting of the horizontal winds. This we justified by the argument that the horizontal symmetry can be broken by the large fluctuations in vertical winds. These horizontal winds are between one to two orders of magnitude larger than the vertical winds in almost all situations.

The observed large vertical winds must have some extent in the horizontal. In studies of auroral arcs, vertical winds have been correlated over a distance of 300 km \cite{article}. Fluctuation in the direction of the vertical winds, upward or downward, at these altitudes was also shown to correspond with the position of the auroral arc. They found in general that upward vertical winds were poleward of the auroral arc and downward winds equatorward of the auroral arc. In the horizontal, the neutral winds have been observed to change direction when auroral arcs appear such that they flow parallel in the region less than 50 km from the arc \cite{horiz_winds_50km_arc}. In \cite{article} the effect of the horizontal winds on the vertical wind structure is discussed (paragraph 30) in a similar vain to the basic estimates we make here. They note that an air parcel rising at \SI{30}{m/s} will take 50 minutes to move from 140 to 240 km (these are the rough altitudes at which they correlated the vertical winds). However the horizontal winds over this altitude range have magnitudes in the range 200 - 500 m/s. They hypothesise that when the horizontal wind is weaker, there may be higher correlation in the vertical wind in the horizontal and or between these two different altitudes. Such discussions and analysis are effectively localised applications of the more general symmetry arguments we made at the beginning of this paper.

If we take the case of a lower magnitude horizontal wind, 200 m/s then over the 20 minutes for a test particle to climb 8.4 km it will have moved $\sim$240 km in the horizontal which is within the maximum correlation scale observed. If the horizontal wind magnitude is just 50 m/s more then we reach this maximum observed 300 km correlation length. However, it could be argued that the correlation measurements are representative of the larger auroral arc which would extend much further than 300 km. Observations of the horizontal winds and modelling do indicate that auroral arcs could form coherent neutral winds in the E emission airglow band in both the horizontal and vertical. More observational studies are required to determine these coherent properties.

There are many examples in the literature where we have sustained periods of vertical winds in the lower thermosphere which may or may not be related to auroral arcs (e.g. \cite{chau2021}). The mechanisms for many of these large vertical winds remain unknown \cite{Larsen2012}. Finally let us take these estimates to the extreme by considering a test particle with dimensions like that of the H1N1 virus which has mass of around 0.8 \si{fg} and diameter of 109 \si{nm}, giving it a much lower density than bacteria of $\sim$ \SI{196}{kgm^{-3}} \cite{Wang16028}. Note that for our disk model we assume that the diameter is equal to the height of the disk. The threshold velocity profile for such a test particle is plotted in Figure \ref{fig:terminal_vel} as the blue dotted curve. Then consider the case where this test particle is caught in a vertical wind of 50 m/s along a long auroral arc. If we assume that the air density the test particle experiences between 110 - 120 km is constant at $\sim 9\times10^{-8}$ \si{kgm^{-3}} as it moves with the updraft, then the test particle will moves upwards with a relative velocity of $\sim$ 2.5 m/s. At this speed for one hour, the test particle will have moved up 9 km.

Vertical winds have been observed to remain upward in a sustained manner for over an hour during geomagnetic storms, so if we imagine the test particle were caught in the upward draft flowing along an auroral arc then it could be carried over 10 km upwards. The height it reaches simply depends on two properties (i) the magnitude of the vertical wind relative to the air density and (ii) how long the Lagrangian trajectory of the test particle remains upward with the necessary threshold velocity. By vertical winds alone we argue it is possible for a virus sized test particle to climb to 120 km. However, we additionally suggest that altitudes greater than 120 km could be possible when multiple climbing factors (e.g. electrostatic levitation, photophoretic forces and vertical winds) act on a single test particle. We also note that there still remains a lot of uncertainty surrounding the large vertical winds in the 80 - 120 km altitude range, primarily due to a lack of consistent temporal and spatial observations, so perhaps even stronger winds are possible via extreme Joule heating. Once a particle has been lofted in an extreme vertical wind event it will experience significant horizontal transport, and will fall unless sustained by sufficient vertical wind or some other force such as those previously identified.

\subsection{\label{sec:level2}Considering downward winds – random walks}

The problem with the current observations is that they give a limited slice of the Eulerian flow in the thermosphere. We believe our undeniably simple approach is valid assuming these large updrafts form mesoscale Lagrangian Coherent structures in which a test particle can be propelled upward. These kinds of observational issues have been tackled with similar approaches in the troposphere for the transport of large dust particles over long distances \cite{long_dist_dust}. In reality the particle's path could be argued to follow some form of random walk in the auroral region, modulated by the solar activity, subsequent aurora and strength of the horizontal winds. The random walk approach may have considerable traction if more observations become available to establish patterns of these stronger vertical winds breaking the natural horizontal circulation symmetry.

In the horizontal we have a reasonably continuous flow which is well documented and modelled. In the vertical the instantaneous large random winds appear as fluctuations, sometimes downward and sometimes upward or in our case sometimes not strong enough to support the weight of the particle. If we consider the Lagrangian motion of a particle, it will follow the Lagrangian Coherent Structure in the horizontal which has been studied before \cite{LCS}. The particle will move across above the surface of the Earth being pushed upward and downward by the local physics (e.g. Joule heating). For each moment of displacement in the horizontal direction it will either move up or down in the vertical.

We considered framing the problem this way as it allows us to ignore the issue of the exact trajectory of the particle and leaves us with the classic statistical physics problem of the random walk in the vertical. This kind of simple application of stochastic physics is not new in atmospheric physics and has been successfully implemented for problems at the surface. However a limiting factor is the air density at these high altitudes which is too low to allow this mean field theoretic approach as the threshold velocity condition is not met for the average or root-mean-square vertical velocities which are between 10-20 m/s \cite{Larsen2012a}. The approach we outlined may become useful as more field campaigns are conducted in the lower thermosphere especially if auroral sub storms can be better observed. Auroral substorms may exhibit vertical and horizontal wind correlations more conducive to this random walk method for particle transport.

Latitudes other than polar ones may well be important for transport via strong vertical winds but the evidence basis for strong vertical winds outside of these regions is weaker, and the auroral arcs provide a ready-made Lagrangian Coherent Structure which enables our 1-D model to capture the basic picture of the dynamics reasonably. At lower latitudes, horizontal dynamics will be far more dominant due to the lack of aurora. Only by further observation field campaigns of both high and low latitudes will it be possible to determine which is dominant for vertical transport.

There is also the question as to how long such large particles would remain in the atmosphere. It has been argued that strong turbulence could keep particles in suspension for a longer time than is expected in the troposphere \cite{garcia}. We suggest that in our case, the mesopause would act as a buffer for particles above it due to the high turbulence in that region from breaking gravity waves and dynamic instabilities \cite{VARNEY2015436}. However, our interest is simply in how high an altitude a particles can reach. Since there exists a small but measurable concentration of micrometer sized particles lower down in the mesosphere at any given time, it should imply that there will be some even smaller concentration at even higher altitudes. 

\section{Discussion}

The purpose of this paper is to show using some simple estimates that particles larger than the air molecules, can be lifted in the upper atmosphere, raising the possibility for biological particles to be projected to higher altitudes than presently have been investigated. The estimates we present are just crude conjecture to an extent. However, it is all that can be estimated given the lack of observations and today's insufficient model capabilities available without resorting to model forcing. Nonetheless, this result has importance; we find that there is the possibility of these larger particles being carried from the upper mesosphere into the thermosphere. Our simple equation for the steady state velocity of a particle blown upwards by the wind can be used by others to determine order of magnitude effects of vertical winds.

By showing that it is possible for large, heavy particles to reach these high altitudes, simply by vertical wind transport, interesting possibilities and questions are opened up. For example, it has been suggested that if biological particles can be found at a minimum altitude of 150 km, then hypervelocity space dust, which continuously impacts the atmosphere, has enough momentum to facilitate the planetary escape of such particles \cite{Berera2017}. It was this work that also suggested vertical winds as a possible mechanism to facilitate the upward climb. Our estimates using reported observations of large vertical winds shows that it is conceivable for such particles to be projected from near the highest measured altitudes in the mesosphere up to 120 km, which is 30 km off the minimum altitude given in the above reference.

The highest altitude that biological particles have been measured is 77 km \cite{Imshenetsky1978}. Using specially adapted meteorological rockets Imshenetsky et al. found bacterial and fungal organisms in the mesosphere between 48 and 77 km. To our knowledge no further studies like this have been conducted since to push this biosphere boundary further. The recent analysis of swabs taken from the exterior of the ISS, which has an altitude of 400 km, does strongly indicate that particles of a biological origin (whole bacteria or DNA fragments) can reach deep into the thermosphere \cite{ISS_DNA}. Our estimates support the hypothesis that these results are from the Earth's atmosphere. Most likely in the form of DNA or organelle fragments due to the size/mass constraints as the higher such particles can be projected into the thermosphere, the more they will be effected by the ascending branch of the global electric circuit as \cite{ISS_DNA} suggested.

So we would suggest that biological particles can be found at altitudes higher than 77 km, especially considering the long history of large vertical winds measured in the upper mesosphere and thermosphere that occur during geomagnetic storms. The results of this paper contribute to the small but building body of evidence that the upper atmosphere should be of interest to the astrobiology community and that further experiment campaigns are needed. We suggest that searches for biological material could be conducted in the mesosphere and potentially the lower thermosphere.

Future work could consider the atmosphere of Mars, for which this vertical winds mechanism might be more suitable due to the frequency of dust storms and the size of the dust. The weight of a particle on Mars is 38\% less than on Earth and the air density at the surface is only 1\%. Dust storms have strong vertical winds and dust particles with radii larger than a micrometre \cite{mars_dusty} \cite{dust_size_mars1}. Vertical particle transport in the Martian atmosphere is therefore likely to be strongly impacted by these storms which can project dust up to around 80 km \cite{mars_dusty}. Smaller particles could attach to the dust and therefore reach higher altitudes. Given that micrometre sized particles can be found at high altitudes around dust storms, then its possible that biological particles of the same size could also be lofted to the same heights. If biological particles are present on Mars, and if such particles can be found at these higher altitudes then they could be sampled and studied by satellites or probes such as balloons, without having to land on the planets surface, using a methodology similar to that of \cite{ISS_DNA} and \cite{satellite_mars_18}.

Our results are complementary to the ideas associated with gravito-photophoresis, which has applications for climate science, as it has been suggested that particles could be engineered to reflect sunlight and be propelled by photophoretic forces \cite{Keith2010}. Future work could examine combinations of thermospheric vertical winds and propulsion by photophoretic force to determine the possible effects of geoengineering in the upper atmosphere. We think it is important to understand the possible impacts of any proposed geoengineering solutions on the upper thermosphere. This region has much more diverse chemical processes catalysed by the stronger solar radiation and ionospheric phenomena. So even if an engineered particle was to reach these high altitudes for a brief period that could be long enough to create unexpected free radicals which over gradual build up could have unfortunate consequences for atmospheric composition, analogous to Ozone depletion due to chlorofluorocarbons \cite{ozone}.

For the middle atmosphere community, the transport of particles into the upper mesosphere and lower thermosphere are usually related to the formation of the cold summer mesopause and noctilucent clouds which are seasonal phenomena, as well as sudden stratospheric warmings (SSWs) which are intraseasonal events \cite{randall2006}\cite{randall2009}\cite{meraner}. Our simple 1-D model neglects cloud microphysics processes such as the charging of the particles, and their ability to be nucleation sites for mesospheric ice particles with subsequent formation of noctilucent clouds. One of the least understood aspects of the formation of the clouds is the nucleation rate \cite{wilms}. One of the key components of this parameter is the vertical wind and therefore the strong vertical winds will have an impact on the nucleation and sedimentation process. Stronger vertical winds will loft more particles to act as nuclei but they will also limit the growth time of the ice particles. Understanding the nucleation rate will greatly assist in setting bounds on transport mechanisms from the stratosphere into the thermosphere. These may have applications to long-term climate modelling \cite{lubken}.

\section{Conclusion}
The observations of large vertical winds reported for decades in the upper mesosphere and thermosphere have been used as the basis for developing a one dimensional vertical transport model for particles larger and heavier  than the air itself. Our model contains two basic ingredients. First, it has the typical forces created by vertical winds. Second, it recognizes a symmetry that the profile of the vertical winds is basically similar over differing horizontal positions over the Earth. This symmetry thus allows the complicated horizontal dynamics to be factored out at the zeroth approximation, leaving only the one-dimensional vertical motion. Exploiting this symmetry alongside the observed forces of these vertical winds, makes this a very strong argument that our model does capture the essential features of upward climb of particles by vertical winds. Any future more developed differential equation to that in this paper or any computer simulations trying to examine this problem in essence would need to capture the basic physics contained in our argument. These potential future developments could provide a more detailed portrayal of the vertical climb of test particles, but the basic features would be what we have argued here.

In the context of other observations and modelling studies of auroral arcs, our estimates indicate that such particles could indeed be transported deeper into the thermosphere than the previously measured and considered value of 77 km. We argue the case that a nanometer sized particle could climb to an altitude of 120 km via vertical winds generated along auroral arcs. We call for further field campaigns to determine better the horizontal distribution of these large vertical winds and particularly between 90 - 150 km, as well as for modelling groups to consider examining three-dimensional Lagrangian coherent structures in the thermosphere.

\section*{Acknowledgment}
AB thanks Jorgen Frederiksen for helpful discussions. DJB acknowledges useful discussion about the representation of vertical winds in non-hydrostatic thermosphere models with Dr.\, Jia Yue and Dr.\, Katherine Garcia-Sage of the NASA Goddard Space Flight Center. AB is partially supported by STFC. DJB is supported by the Carnegie Trust and STFC. We thank the two referees for their careful and insightful review of our manuscript.

\bibliographystyle{unsrt}
\bibliography{refs}

\begin{thebibliography}{10}

\bibitem{SMITH19981425}
Roger~W. Smith.
\newblock Vertical winds: a tutorial.
\newblock {\em Journal of Atmospheric and Solar-Terrestrial Physics},
  60(14):1425 -- 1434, 1998.

\bibitem{Larsen2012}
M.~F. Larsen and J.~W. Meriwether.
\newblock {Vertical winds in the thermosphere}.
\newblock {\em Journal of Geophysical Research: Space Physics}, 117(9):1--10,
  2012.

\bibitem{Rees1984}
D~Rees, R~W Smith, P~J Charleton, F~G McCormac, N~Lloyd, and {\AA}ke Steen.
\newblock {The generation of vertical thermospheric winds and gravity waves at
  auroral latitudes—I. Observations of vertical winds}.
\newblock {\em Planetary and Space Science}, 32(6):667--684, 1984.

\bibitem{Price1995}
G~D Price, R~W Smith, and G~Hernandez.
\newblock {Simultaneous measurements of large vertical winds in the upper and
  lower thermosphere}.
\newblock {\em Journal of Atmospheric and Terrestrial Physics}, 57(6):631--643,
  1995.

\bibitem{Guo2003}
W~Guo and D~J McEwen.
\newblock {Vertical winds in the central polar cap}.
\newblock {\em Geophysical Research Letters}, 30(14), 2003.

\bibitem{Wardill1986}
P~Wardill and F~Jacka.
\newblock {Vertical motions in the thermosphere over Mawson, Antarctica}.
\newblock {\em Journal of Atmospheric and Terrestrial Physics}, 48(3):289--292,
  1986.

\bibitem{Crickmore1993}
R~I Crickmore.
\newblock {A comparison between vertical winds and divergence in the
  high-latitude thermosphere}.
\newblock {\em Annales Geophysicae}, 11:728--733, jan 1993.

\bibitem{Crickmore1991}
R~I Crickmore, J~R Dudeney, and A~S Rodger.
\newblock {Vertical thermospheric winds at the equatorward edge of the auroral
  oval}.
\newblock {\em Journal of Atmospheric and Terrestrial Physics}, 53(6):485--492,
  1991.

\bibitem{Aruliah1995}
Anasuya~L Aruliah and David Rees.
\newblock {The trouble with thermospheric vertical winds: geomagnetic, seasonal
  and solar cycle dependence at high latitudes}.
\newblock {\em Journal of Atmospheric and Terrestrial Physics}, 57(6):597--609,
  1995.

\bibitem{Ishii2001}
M~Ishii, M~Conde, R~W Smith, M~Krynicki, E~Sagawa, and S~Watari.
\newblock {Vertical wind observations with two Fabry-Perot interferometers at
  Poker Flat, Alaska}.
\newblock {\em Journal of Geophysical Research: Space Physics},
  106(A6):10537--10551, jun 2001.

\bibitem{Greet2002}
P~A Greet, J~L Innis, and P~L Dyson.
\newblock {Thermospheric vertical winds in the auroral oval/polar cap region}.
\newblock {\em Ann. Geophys.}, 20(12):1987--2001, dec 2002.

\bibitem{CONDE1995589}
M.~Conde and P.L. Dyson.
\newblock Thermospheric vertical winds above mawson, antarctica.
\newblock {\em Journal of Atmospheric and Terrestrial Physics}, 57(6):589 --
  596, 1995.

\bibitem{Anderson2012}
C~Anderson, M~Conde, and M~G McHarg.
\newblock {Neutral thermospheric dynamics observed with two scanning Doppler
  imagers: 2. Vertical winds}.
\newblock {\em Journal of Geophysical Research: Space Physics}, 117(A3), 2012.

\bibitem{ronksley}
A.~Ronksley.
\newblock {\em Optical Remote Sensing of Mesoscale Thermospheric Dynamics Above
  Svalbard and Kiruna}.
\newblock PhD thesis, University College London, 2016.

\bibitem{Oyama}
S.~Oyama, B.~J. Watkins, S.~Nozawa, S.~Maeda, and M.~Conde.
\newblock Vertical ion motion observed with incoherent scatter radars in the
  polar lower ionosphere.
\newblock {\em Journal of Geophysical Research: Space Physics}, 110(A4), 2005.

\bibitem{Anderson2011}
C.~Anderson, T.~Davies, M.~Conde, P.~Dyson, and M.~J. Kosch.
\newblock {Spatial sampling of the thermospheric vertical wind field at auroral
  latitudes}.
\newblock {\em Journal of Geophysical Research: Space Physics}, 116(6):1--14,
  2011.

\bibitem{chau2021}
J.~L. Chau, R.~Marino, F.~Feraco, J.~M. Urco, G.~Baumgarten, F.-J. Lübken,
  W.~K. Hocking, C.~Schult, T.~Renkwitz, and R.~Latteck.
\newblock Radar observation of extreme vertical drafts in the polar summer
  mesosphere.
\newblock {\em Geophysical Research Letters}, 48(16):e2021GL094918, 2021.
\newblock e2021GL094918 2021GL094918.

\bibitem{chu2021}
Xinzhao Chu, Yingfei Chen, Chihoko~Y. Cullens, Zhibin Yu, Zhonghua Xu,
  Shun-Rong Zhang, Wentao Huang, Jackson Jandreau, Thomas~J. Immel, and
  Arthur~D. Richmond.
\newblock Mid-latitude thermosphere-ionosphere na (tina) layers observed with
  high-sensitivity na doppler lidar over boulder (40.13°n, 105.24°w).
\newblock {\em Geophysical Research Letters}, 48(11):e2021GL093729, 2021.
\newblock e2021GL093729 2021GL093729.

\bibitem{Griffin2018}
Daniel~Joe Griffin.
\newblock {\em {The Extension of a Non-Hydrostatic Dynamical Core into the
  Thermosphere}}.
\newblock PhD thesis, University of Exeter, 2018.

\bibitem{Priv_Comm_kath}
Katherine Garcia-Sage.
\newblock {Private Communication}, 2020.
\newblock NASA Goddard Space Flight Center.

\bibitem{Deng2008}
Yue Deng, Arthur~D Richmond, Aaron~J Ridley, and Han-Li Liu.
\newblock {Assessment of the non-hydrostatic effect on the upper atmosphere
  using a general circulation model (GCM)}.
\newblock {\em Geophysical Research Letters}, 35(1), 2008.

\bibitem{Zhu2017}
Qingyu Zhu, Yue Deng, Astrid Maute, Cheng Sheng, and Cissi~Y. Lin.
\newblock {Impact of the vertical dynamics on the thermosphere at low and
  middle latitudes: GITM simulations}.
\newblock {\em Journal of Geophysical Research: Space Physics},
  122(6):6882--6891, 2017.

\bibitem{Zhu2020}
Qingyu Zhu, Yue Deng, Arthur Richmond, Astrid Maute, Yun-Ju Chen, Marc
  Hairston, Liam Kilcommons, Delores Knipp, Robert Redmon, and Elizabeth
  Mitchell.
\newblock {Impacts of Binning Methods on High-Latitude Electrodynamic Forcing:
  Static Versus Boundary-Oriented Binning Methods}.
\newblock {\em Journal of Geophysical Research: Space Physics},
  125(1):e2019JA027270, 2020.

\bibitem{Ridley2006}
A~J Ridley, Y~Deng, and G~T{\'{o}}th.
\newblock {The global ionosphere–thermosphere model}.
\newblock {\em Journal of Atmospheric and Solar-Terrestrial Physics},
  68(8):839--864, 2006.

\bibitem{Priv_Comm_Jia}
Jia Yue.
\newblock {Private Communication}, 2020.
\newblock NASA Goddard Space Flight Center.

\bibitem{Heintzenberg}
J.~Heintzenberg, M.~Hermann, and D.~Theiss.
\newblock Out of africa: High aerosol concentrations in the upper troposphere
  over africa.
\newblock {\em Atmospheric Chemistry and Physics}, 3(4):1191--1198, 2003.

\bibitem{rosen}
James~M. Rosen.
\newblock The vertical distribution of dust to 30 kilometers.
\newblock {\em Journal of Geophysical Research (1896-1977)}, 69(21):4673--4676,
  1964.

\bibitem{ursem}
Bob Ursem.
\newblock Climate shifts and the role of nano structured particles in the
  atmosphere.
\newblock {\em Atmospheric and Climate Sciences}, 06:51--76, 01 2016.

\bibitem{XU2003201}
Li~Xu, Guangyu Shi, Li~Zhang, Jun Zhou, and Yasunobu Iwasaka.
\newblock Number concentration, size distribution and fine particle fraction of
  tropospheric and stratospheric aerosols.
\newblock {\em China Particuology}, 1(5):201 -- 205, 2003.

\bibitem{yinyan}
Yan Yin, Qian Chen, Lianji Jin, Baojun Chen, Shichao Zhu, and Xiaopei Zhang.
\newblock The effects of deep convection on the concentration and size
  distribution of aerosol particles within the upper troposphere: A case study.
\newblock {\em Journal of Geophysical Research (Atmospheres)}, 117:22202--, 11
  2012.

\bibitem{Griffin2004}
Dale Griffin.
\newblock {Terrestrial microorganisms at 20,000 meters in Earth's atmosphere}.
\newblock {\em Aerobiologia}, 20:135--140, jan 2004.

\bibitem{Wainwright2003}
M~Wainwright, N~C Wickramasinghe, J~V Narlikar, and P~Rajaratnam.
\newblock {Microorganisms cultured from stratospheric air samples obtained at
  41 km.}
\newblock {\em FEMS microbiology letters}, 218(1):161--165, jan 2003.

\bibitem{Imshenetsky1978}
A~A Imshenetsky, S~V Lysenko, and G~A Kazakov.
\newblock {Upper boundary of the biosphere}.
\newblock {\em Applied and environmental microbiology}, 35(1):1--5, jan 1978.

\bibitem{ISS_DNA}
T.~V. Grebennikova, A.~V. Syroeshkin, E.~V. Shubralova, O.~V. Eliseeva, L.~V.
  Kostina, N.~Y. Kulikova, O.~E. Latyshev, M.~A. Morozova, A.~G. Yuzhakov,
  I.~A. Zlatskiy, and et~al.
\newblock The dna of bacteria of the world ocean and the earth in cosmic dust
  at the international space station.
\newblock {\em The Scientific World Journal}, page 1–7, Apr 2018.

\bibitem{Rohatschek1996}
Hans Rohatschek.
\newblock {Levitation of stratospheric and mesospheric aerosols by
  gravito-photophoresis}.
\newblock {\em Journal of Aerosol Science}, 27(3):467--475, 1996.

\bibitem{Horvath2014}
Helmuth Horvath.
\newblock {Photophoresis – a Forgotten Force?}
\newblock {\em KONA Powder and Particle Journal}, 31:181--199, 2014.

\bibitem{Wallace}
John~M. Wallace and Peter~V. Hobbs.
\newblock {\em {Atmospheric Science: An Introductory Survey - John M. Wallace,
  Peter V. Hobbs}}.
\newblock Academic Press, 1977.

\bibitem{2006JASTP..68..715R}
Markus {Rapp} and Gary~E. {Thomas}.
\newblock {Modeling the microphysics of mesospheric ice particles: Assessment
  of current capabilities and basic sensitivities}.
\newblock {\em Journal of Atmospheric and Solar-Terrestrial Physics},
  68(7):715--744, April 2006.

\bibitem{plane2012}
John M.~C. Plane.
\newblock Cosmic dust in the earth{'}s atmosphere.
\newblock {\em Chem. Soc. Rev.}, 41:6507--6518, 2012.

\bibitem{wilms}
Henrike Wilms, Markus Rapp, and Annekatrin Kirsch.
\newblock Nucleation of mesospheric cloud particles: Sensitivities and limits.
\newblock {\em Journal of Geophysical Research: Space Physics},
  121(3):2621--2644, 2016.

\bibitem{tupper}
Andrew Tupper, Christiane Textor, Michael Herzog, Hans-F. Graf, and Michael
  Richards.
\newblock {Tall clouds from small eruptions: the sensitivity of eruption height
  and fine ash content to tropospheric instability}.
\newblock {\em Natural Hazards: Journal of the International Society for the
  Prevention and Mitigation of Natural Hazards}, 51(2):375--401, November 2009.

\bibitem{wilson}
L.~Wilson, R.~S.~J. Sparks, T.~C. Huang, and N.~D. Watkins.
\newblock The control of volcanic column heights by eruption energetics and
  dynamics.
\newblock {\em Journal of Geophysical Research: Solid Earth},
  83(B4):1829--1836, 1978.

\bibitem{self}
S.~Self and M.~R. Rampino.
\newblock The 1883 eruption of krakatau.
\newblock {\em Nature}, 294:699--704, 1981.

\bibitem{VERBEEK1884}
R~D~M Verbeek.
\newblock {The Krakatoa Eruption1}.
\newblock {\em Nature}, 30(757):10--15, 1884.

\bibitem{ludlam}
F.~H. Ludlam.
\newblock Noctilucent clouds.
\newblock {\em Tellus}, 9(3):341--364, 1957.

\bibitem{electrostatic_volcano}
Matthew~J. Genge.
\newblock {Electrostatic levitation of volcanic ash into the ionosphere and its
  abrupt effect on climate}.
\newblock {\em Geology}, 46(10):835--838, 08 2018.

\bibitem{WEBBER2013138}
Derek Webber.
\newblock Space tourism: Its history, future and importance.
\newblock {\em Acta Astronautica}, 92(2):138--143, 2013.
\newblock 2nd IAA Symposium on Private Human Access to Space.

\bibitem{astrobio_smith}
David Smith.
\newblock Microbes in the upper atmosphere and unique opportunities for
  astrobiology research.
\newblock {\em Astrobiology}, 13, 10 2013.

\bibitem{dassarma_antunes_dassarma_2020}
Priya Dassarma, André Antunes, Marta~Filipa Simões, and Shiladitya Dassarma.
\newblock Earths stratosphere and microbial life.
\newblock {\em Astrobiology: Current, Evolving, and Emerging Perspectives}, Jan
  2020.

\bibitem{Greaves2020}
Jane~S. Greaves, Anita M.~S. Richards, William Bains, Paul~B. Rimmer, Hideo
  Sagawa, David~L. Clements, Sara Seager, Janusz~J. Petkowski, Clara
  Sousa-Silva, Sukrit Ranjan, Emily Drabek-Maunder, Helen~J. Fraser, Annabel
  Cartwright, Ingo Mueller-Wodarg, Zhuchang Zhan, Per Friberg, Iain Coulson,
  E'lisa Lee, and Jim Hoge.
\newblock Phosphine gas in the cloud decks of venus.
\newblock {\em Nature Astronomy}, Sep 2020.

\bibitem{venus_debunked}
{Snellen, I. A. G.}, {Guzman-Ramirez, L.}, {Hogerheijde, M. R.}, {Hygate, A. P.
  S.}, and {van der Tak, F. F. S.}
\newblock Re-analysis of the 267 ghz alma observations of venus - no
  statistically significant detection of phosphine.
\newblock {\em A\&A}, 644:L2, 2020.

\bibitem{Hallsworth2021}
John~E. Hallsworth, Thomas Koop, Tiffany~D. Dallas, Mar{\'i}a-Paz Zorzano,
  Juergen Burkhardt, Olga~V. Golyshina, Javier Mart{\'i}n-Torres, Marcus~K.
  Dymond, Philip Ball, and Christopher~P. McKay.
\newblock Water activity in venus's uninhabitable clouds and other planetary
  atmospheres.
\newblock {\em Nature Astronomy}, Jun 2021.

\bibitem{WACCM}
A.~Gettelman, M.~J. Mills, D.~E. Kinnison, R.~R. Garcia, A.~K. Smith, D.~R.
  Marsh, S.~Tilmes, F.~Vitt, C.~G. Bardeen, J.~McInerny, H.-L. Liu, S.~C.
  Solomon, L.~M. Polvani, L.~K. Emmons, J.-F. Lamarque, J.~H. Richter, A.~S.
  Glanville, J.~T. Bacmeister, A.~S. Phillips, R.~B. Neale, I.~R. Simpson,
  A.~K. DuVivier, A.~Hodzic, and W.~J. Randel.
\newblock The whole atmosphere community climate model version 6 (waccm6).
\newblock {\em Journal of Geophysical Research: Atmospheres},
  124(23):12380--12403, 2019.

\bibitem{CMAM}
C.~McLandress, W.~E. Ward, V.~I. Fomichev, K.~Semeniuk, S.~R. Beagley, N.~A.
  McFarlane, and T.~G. Shepherd.
\newblock Large-scale dynamics of the mesosphere and lower thermosphere: An
  analysis using the extended canadian middle atmosphere model.
\newblock {\em Journal of Geophysical Research: Atmospheres}, 111(D17), 2006.

\bibitem{solomon_15}
Stanley~C. Solomon, Liying Qian, and Raymond~G. Roble.
\newblock New 3-d simulations of climate change in the thermosphere.
\newblock {\em Journal of Geophysical Research: Space Physics},
  120(3):2183--2193, 2015.

\bibitem{thermo_climate}
Liying Qian, Christoph Jacobi, and Joseph McInerney.
\newblock Trends and solar irradiance effects in the mesosphere.
\newblock {\em Journal of Geophysical Research: Space Physics},
  124(2):1343--1360, 2019.

\bibitem{thermo_climate_2}
Stanley~C. Solomon, Han-Li Liu, Daniel~R. Marsh, Joseph~M. McInerney, Liying
  Qian, and Francis~M. Vitt.
\newblock Whole atmosphere climate change: Dependence on solar activity.
\newblock {\em Journal of Geophysical Research: Space Physics},
  124(5):3799--3809, 2019.

\bibitem{LCS}
N.~Wang, U.~Ramirez, F.~Flores, and S.~Datta-Barua.
\newblock Lagrangian coherent structures in the thermosphere: Predictive
  transport barriers.
\newblock {\em Geophysical Research Letters}, 44(10):4549--4557, 2017.

\bibitem{article}
M.~Ishii, M.~Kubota, Mark Conde, Roger Smith, and M.~Krynicki.
\newblock Vertical wind distribution in the polar thermosphere during
  horizontal e region experiment (hex) campaign.
\newblock {\em Journal of Geophysical Research}, 109, 12 2004.

\bibitem{Price1991}
G.~D. Price, F.~Jacka, R.~A. Vincent, and G.~B. Burns.
\newblock {The influence of geomagnetic activity on the upper mesosphere lower
  thermosphere in the auroral zone. II. Horizontal winds}.
\newblock {\em Journal of Atmospheric and Terrestrial Physics},
  53(10):923--947, 1991.

\bibitem{rubberbible}
W.M. Haynes.
\newblock {\em {CRC Handbook of Chemistry and Physics, 95th Edition,
  2014-2015}}, volume~54.
\newblock CRC Press, Boca Raton, 95 edition, 2014.

\bibitem{doi:10.1063/1.2393436}
Nicolas~G. Hadjiconstantinou.
\newblock The limits of navier-stokes theory and kinetic extensions for
  describing small-scale gaseous hydrodynamics.
\newblock {\em Physics of Fluids}, 18(11):111301, 2006.

\bibitem{Karniadakis2005}
S.S. Antman, J.E. Marsden, and L.~Sirovich, editors.
\newblock {\em Basic Concepts and Technologies}, pages 1--48.
\newblock Springer New York, New York, NY, 2005.

\bibitem{WHO}
Hazard prevention and control in the work environment: Airborne dust.
  protection of the human environment occupational health and environmental
  health series.
\newblock {\em The Annals of Occupational Hygiene}, 2000.

\bibitem{2018cosp...42E.893D}
Douglas Drob, John Emmert, David Siskind, and J~Michael Picone.
\newblock {NRLMSIS 2.0: New Formulation, New Data}.
\newblock In {\em 42nd COSPAR Scientific Assembly}, volume~42, pages
  C4.2--3--18, jul 2018.

\bibitem{Picone2002}
J~M Picone, A~E Hedin, D~P Drob, and A~C Aikin.
\newblock {NRLMSISE-00 empirical model of the atmosphere: Statistical
  comparisons and scientific issues}.
\newblock {\em Journal of Geophysical Research: Space Physics}, 107(A12):SIA
  15--1--SIA 15--16, 2002.

\bibitem{fluids}
Caleb Bell.
\newblock {fluids: Fluid dynamics component of Chemical Engineering Design
  Library (ChEDL)}, 2020.

\bibitem{Wang16028}
Shaopeng Wang, Xiaonan Shan, Urmez Patel, Xinping Huang, Jin Lu, Jinghong Li,
  and Nongjian Tao.
\newblock Label-free imaging, detection, and mass measurement of single viruses
  by surface plasmon resonance.
\newblock {\em Proceedings of the National Academy of Sciences},
  107(37):16028--16032, 2010.

\bibitem{horiz_winds_50km_arc}
M.~Kosch, C.~Anderson, Roman Makarevich, Brett Carter, R.~Fiori, M.~Conde,
  P.~Dyson, and T.~Davies.
\newblock First e-region observations of meso-scale neutral wind interaction
  with auroral arcs.
\newblock {\em Journal of Geophysical Research: Space Physics}, 115, 02 2010.

\bibitem{long_dist_dust}
Mich{\`e}lle van~der Does, Peter Knippertz, Philipp Zschenderlein,
  R.~Giles~Harrison, and Jan-Berend~W. Stuut.
\newblock The mysterious long-range transport of giant mineral dust particles.
\newblock {\em Science Advances}, 4(12), 2018.

\bibitem{Larsen2012a}
M~F Larsen and J~W Meriwether.
\newblock {Vertical winds in the thermosphere}.
\newblock {\em Journal of Geophysical Research: Space Physics}, 117(A9), sep
  2012.

\bibitem{garcia}
L.~Garcia-Carreras, D.~J. Parker, J.~H. Marsham, P.~D. Rosenberg, I.~M. Brooks,
  A.~P. Lock, F.~Marenco, J.~B. McQuaid, and M.~Hobby.
\newblock {The Turbulent Structure and Diurnal Growth of the Saharan
  Atmospheric Boundary Layer}.
\newblock {\em Journal of the Atmospheric Sciences}, 72(2):693--713, 02 2015.

\bibitem{VARNEY2015436}
R.H. Varney and M.C. Kelley.
\newblock Mesosphere | polar summer mesopause.
\newblock In Gerald~R. North, John Pyle, and Fuqing Zhang, editors, {\em
  Encyclopedia of Atmospheric Sciences (Second Edition)}, pages 436 -- 443.
  Academic Press, Oxford, second edition edition, 2015.

\bibitem{Berera2017}
Arjun Berera.
\newblock {Space Dust Collisions as a Planetary Escape Mechanism}.
\newblock {\em Astrobiology}, 17(12):1274--1282, nov 2017.

\bibitem{mars_dusty}
Nicholas~G. Heavens, David~M. Kass, and James~H. Shirley.
\newblock Dusty deep convection in the mars year 34 planet-encircling dust
  event.
\newblock {\em Journal of Geophysical Research: Planets}, 124(11):2863--2892,
  2019.

\bibitem{dust_size_mars1}
M.~T. Lemmon, S.~D. Guzewich, T.~McConnochie, A.~de~Vicente-Retortillo,
  G.~Martínez, Michael~D. Smith, J.~F. Bell~III, D.~Wellington, and S.~Jacob.
\newblock Large dust aerosol sizes seen during the 2018 martian global dust
  event by the curiosity rover.
\newblock {\em Geophysical Research Letters}, 46(16):9448--9456, 2019.

\bibitem{satellite_mars_18}
Kai {Wickhusen}, J{\"u}rgen {Oberst}, and Friedrich {Damme}.
\newblock {A Proposed Mission to Very Low Mars Orbit - Supported by an Electric
  Propulsion System}.
\newblock In {\em European Planetary Science Congress}, volume~12, pages
  EPSC2018--364, September 2018.

\bibitem{Keith2010}
David~W. Keith.
\newblock {Photophoretic levitation of engineered aerosols for geoengineering}.
\newblock {\em Proceedings of the National Academy of Sciences of the United
  States of America}, 107(38):16428--16431, 2010.

\bibitem{ozone}
Martyn~P. Chipperfield.
\newblock Global atmosphere – the antarctic ozone hole.
\newblock In {\em Still Only One Earth: Progress in the 40 Years Since the
  First UN Conference on the Environment}, pages 1--33. The Royal Society of
  Chemistry, 2015.

\bibitem{randall2006}
C.~E. Randall, V.~L. Harvey, C.~S. Singleton, P.~F. Bernath, C.~D. Boone, and
  J.~U. Kozyra.
\newblock Enhanced nox in 2006 linked to strong upper stratospheric arctic
  vortex.
\newblock {\em Geophysical Research Letters}, 33(18), 2006.

\bibitem{randall2009}
C.~E. Randall, V.~L. Harvey, D.~E. Siskind, J.~France, P.~F. Bernath, C.~D.
  Boone, and K.~A. Walker.
\newblock Nox descent in the arctic middle atmosphere in early 2009.
\newblock {\em Geophysical Research Letters}, 36(18), 2009.

\bibitem{meraner}
Katharina Meraner and Hauke Schmidt.
\newblock Transport of nitrogen oxides through the winter mesopause in
  hammonia.
\newblock {\em Journal of Geophysical Research: Atmospheres},
  121(6):2556--2570, 2016.

\bibitem{lubken}
Franz-Josef Lübken, Uwe Berger, and Gerd Baumgarten.
\newblock On the anthropogenic impact on long-term evolution of noctilucent
  clouds.
\newblock {\em Geophysical Research Letters}, 45(13):6681--6689, 2018.

\end{thebibliography}

\end{document}